\documentclass{article}

\usepackage[margin=3.2cm]{geometry}

\usepackage{mathptmx}
\usepackage[scaled=.90]{helvet}
\usepackage[scale=0.9]{inconsolata}

\usepackage[protrusion=true,expansion=true]{microtype}

\usepackage{parskip}

\usepackage{hyperref}
\usepackage{url}
\usepackage{graphicx}
\usepackage{mathtools}
\usepackage{xspace}
\usepackage{colortbl, booktabs} 
\usepackage{subcaption}
\usepackage{tabularx}
\usepackage{array}
\usepackage{makecell}
\usepackage{enumitem}
\usepackage{overpic}

\usepackage[auth-lg,affil-sl]{authblk}

\author[]{Daphne Ippolito}
\author[]{Ann Yuan}
\author[]{Andy Coenen}
\author[]{Sehmon Burnam}

\affil[]{Google Research}
\affil[]{\texttt{\{dei,annyuan,andycoenen,sehmon\}@google.com}}

\date{}

\usepackage{natbib}

\usepackage[utf8]{inputenc} 
\usepackage[T1]{fontenc}    
\usepackage{hyperref}       
\usepackage{url}            
\usepackage{booktabs}       
\usepackage{amsfonts}       
\usepackage{nicefrac}       
\usepackage{microtype}      
\usepackage{xcolor}         
\usepackage{amsmath,amsthm}

\theoremstyle{plain}

\theoremstyle{definition}

\theoremstyle{remark}

\title{Creative Writing with an AI-Powered Writing Assistant: Perspectives from Professional Writers}

\usepackage{listings}
\newcommand{\q}[1]{``\textit{#1''}}

\definecolor{blue1}{rgb}{0,0,0.1}

\begin{document}
\maketitle

\begin{abstract}
Recent developments in natural language generation (NLG) using neural language models have brought us closer than ever to the goal of building AI-powered creative writing tools.
However, most prior work on human-AI collaboration in the creative writing domain has evaluated new systems with amateur writers, typically in contrived user studies of limited scope.
In this work, we commissioned 13 professional, published writers from a diverse set of creative writing backgrounds to craft stories using Wordcraft, a text editor with built-in AI-powered writing assistance tools.
Using interviews and participant journals,
we discuss the potential of NLG to have significant impact in the creative writing domain--especially with respect to brainstorming, generation of story details, world-building, and research assistance.
Experienced writers, more so than amateurs, typically have well-developed systems and methodologies for writing, as well as distinctive voices and target audiences.
Our work highlights the challenges in building for these writers; NLG technologies struggle to preserve style and authorial voice, and they lack deep understanding of story contents.
In order for AI-powered writing assistants to realize their full potential, it is essential that they take into account the diverse goals and expertise of human writers.
\end{abstract}

\section{Introduction}

Writing complete stories is considered a hallmark display of human intelligence, and thus researchers in artificial intelligence (AI) and natural language generation (NLG) have long used it as a pinnacle task for their research \citep{klein1973automatic,meehan1977tale,Turner:1993:MCM:166478,dehn1981story,liu2002makebelieve,mcintyre2009learning}.
Creative writing and storytelling present unique challenges for automatic language generation: story arcs extend over thousands of words, stories typically contain multiple characters with their own distinctive personas and voices, and well-written stories have an authorial voice that is consistent and identifiable.
At the same time, lies and fabrications--common generation flaws which are a liability in tasks like machine translation and automatic summarization--can be an asset in the creative domain.

In recent years, the field of NLG has progressed by leaps and bounds due to the development of neural language models
capable of learning the structure of language by ingesting billions of written words
\citep{chowdhery2022palm,zhang2022opt,brown2020language}. 
There has been considerable work in applying these advancements toward the development of AI-powered tools for creative writing, but nearly all previous research in this space has evaluated their methods either with amateur writers or with crowd workers paid to assess performance on narrowly defined tasks \citep{clark2018creative,roemmele2015creative,nichols2020collaborative}.
While these sorts of evaluations are valuable as preliminary assessments, we believe it is also crucial to solicit feedback from actual domain experts in creative writing: professional writers, educators, and language experts.
Skilled writers comprise a unique user group with a different set of needs and expectations than amateurs.
They may be better equipped than amateurs to assess where AI-powered creative writing technology can fit into writing workflows, and how this technology needs to improve to become more impactful and useful.
For example, they are more likely to be sensitive to the issue of voice (and have their own unique voices), and they are more likely to have already honed writing processes.

\begin{figure}[t]
    \centering
    \includegraphics[width=1.05\textwidth]{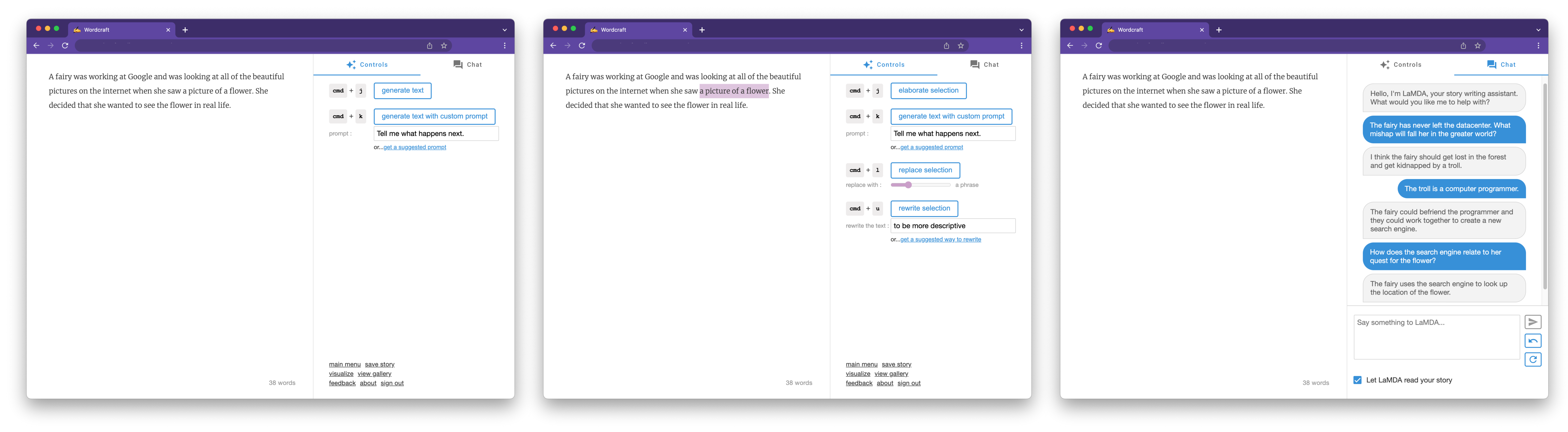}
    \caption{Screenshots of the Wordcraft user interface when no text is selected (left), when text is selected (middle), and in the chatbot interface (right).}
    \label{fig:wordcraft}
\end{figure}

In this paper, we aim to better understand how professional writers perceive and interact with state-of-the-art language generation tools for creative writing.
To accomplish this, we commissioned 13 (1) \textbf{published} writers from (2) \textbf{diverse} backgrounds to use an NLG-powered text editor over an (3) \textbf{extended period of time}, collecting open-ended qualitative feedback along the way.
To the best of our knowledge, this work is also the first to evaluate NLG applications for creative writing with a diverse audience both in terms of background (including participants from a variety of countries and ethnicities) and in terms of expertise (including scriptwriters, poets, educators, novelists, and more).
Our study is also novel in its format; rather than asking participants to adopt a contrived workflow, we invited them to incorporate the provided tools into their writing process in whatever way they saw fit.
Participants were engaged over a period of eight weeks, in line with industry norms for commissioned works of fiction.
This long duration allowed participants to become intimately familiar with the capabilities of the provided tools in order to provide rich feedback regarding their capabilities and deficiencies.

Participants were given access to a tool called Wordcraft \citep{yuan2022wordcraft}, a website consisting of a standard text processor with various additional controls users could harness to generate from or modify the text they had written so far, as well as a conversational chatbot interface.
Both the controls and the chatbot were supported by LaMDA, a large neural language model trained on public dialog data and other public web documents \citep{thoppilan2022lamda}.

From written feedback and conversations with participants, we learned about the workflows for which Wordcraft worked well, and where there is still room for improvement.
Participants desired a tool that was variously a brainstorming partner, a co-writer, a beta reader, and a research assistant.
Some participants cared most about it having the ability to produce high-level plot and narrative ideas while others wanted it to be capable of producing phrases and passages which were good enough to be pasted directly into a story.
Participants emphasized that the user interface of the tool matters as much as the underlying language model backing it.


Participants also spoke extensively about the limitations of the technology--that the generations lacked a distinctive voice and the suggestions were uninteresting, and that it was difficult to control the tool to accomplish specific writing tasks.
The tool's bland suggestions posed an important dilemma.
A system that always errs on the side of avoiding transgression is hamstringing itself from ever achieving human-level creativity, which is often grounded in a rejection of tropes and norms.

In this whitepaper, we first describe the expectations participants had before starting the project.
We then describe the workflows participants experimented with and the limitations participants encountered.
We end with a discussion of lessons learned--areas where technology built with large language models needs to improve before it can achieve broad usefulness in the creative domain and beyond.


\begin{table}[t]
    \centering
    \small
    \caption{Controls implemented in Wordcraft}
    \label{tab:controls}
    \begin{tabular}{p{1in}|p{1.7in}|p{2.5in}}
         \toprule
         \textbf{Control} & \textbf{What does it do?} & \textbf{Example} \\
         \midrule
         Story seeds &
         When the editor is empty, the AI can propose story ideas according to a user-specified prompt,
         &
         A user requests ``a folktale about a fairy working at Google.'' Wordcraft suggests ``A fairy was working at Google and was looking at all of the beautiful pictures on the internet when she saw a picture of a flower. She decided that she wanted to see the flower in real life.''
         \\
         \midrule
         Continuation
         &
         The AI proposes a continuation to follow the text to the left of the user's cursor.
         &
         A story starts with ``The fairy fluttered from rack to rack, peering into each hard drive.'' Wordcraft suggests the continuation: ``The fairy had been sent to retrieve the last surviving copy of the software, and she was determined to do her job.''
         \\
         \midrule
         Elaborate selection
         &
         The text the user has selected is expanded upon. 
         &
         The user selected the word ``software'' in the previous example. Wordrcaft suggests: ``It was a very old program from the 1990s, it was a game called `Oregon Trail'.''
        \\
        \midrule
        Generate text from a custom prompt
        &
        The user can type in any prompt to the model, for example ``tell me what happens next".
        &
        The user types ``tell me more about the fairy's day job.'' Wordcraft suggests ``She was a datacenter fairy, and she helped keep the internet running.''
        \\
        \midrule
        Replace selection
        &
        A fill-in-the-blank control. The user selects text and sees suggestions for alternative text which could fit in that position.
        &
        The user selects the phrase ``whirring of fans'' in the passage: ``The datacenter was alive with the whirring of fans. It did not daunt the fairy.'' Wordrcaft suggests an alternative phrase: ``sounds of the internet.''
        \\
        \hline
        Rewrite selection
        &
        The user can ask Wordcraft to rewrite their selected text according to some property. For example, they could ask it to write the text "to be Shakepearean" or "to include a metaphor."
        &
        The user asks for the sentence ``The fairy was sad.'' to be rewritten to be ``more melodramatic.'' Wordcraft suggests: ``The fairy was broken-hearted, as she knew she would never be able to see the beautiful flower in real life.''
    \end{tabular}
\end{table}

\section{Related Work}
The idea of a computer program that can generate parts of a story or even a story in its entirety has been of interest to computer science researchers since the field's inception.
Early work in this area relied on classical AI algorithms, such as symbolic and logical planning and graph traversal, to generate stories, typically with some level of user control such as being able to specify an initial setting of goals and conditions \citep{klein1973automatic,meehan1977tale,Turner:1993:MCM:166478,dehn1981story}.
More recent work took a data-driven approach, consulting a knowledge base of commonsense assertions or world facts to improve story coherence \citep{liu2002makebelieve,mcintyre2009learning}.
Among recent work, some have strived to generate entire stories without human intervention \citep{fan2018hierarchical} while others have stressed the importance of designing AI systems that prioritize human involvement in the story creation process \citep{riedl2006linear,swanson2021story,akoury2020storium,roemmele2015creative}. 
With Wordcraft, we take the latter position, focusing on the potential of story writing AI to be useful \textit{tool} for writers, just as a traditional word processor, a spellchecker, or a deck of ideation cards are tools to expedite the writing process. 

In recent years, large neural language models--capable of generating natural-sounding text given a natural language prompt--have been applied widely to the development of creative writing tools.
Many of these systems adopt the paradigm of the user and the language model taking turns to append content to the end of the story \citep{nichols2020collaborative,calderwood2020novelists}.
This is the most intuitive paradigm for language models, but it is not necessarily the most intuitive paradigm for human writers.
In Wordcraft, we aim to support a variety of story editing operations, in addition to suggesting story continuations.

Most past user studies involving these story-writing AI have conducted evaluation in contrived settings testing narrow functionality, typically with amateur writers \citep{clark2018creative,yuan2022wordcraft,roemmele2015creative,roemmele2021inspiration,nichols2020collaborative}.
One notable exception is the work of \citet{akoury2020storium}, who incorporated a suggestion engine into an online story writing game and analyzed how game users interacted with it.
Perhaps closest to our work are that of \citet{mirowski2022co}, who hired expert playwrights to cowrite scripts using a language model that suggested characters, scene summaries, and other script components, and that of \citet{calderwood2020novelists}, who observed four professional novelists experimenting with GPT-2.
However, the settings of these works were still quite limited; in both, writers had under two hours to interact with the systems.
In contrast, we gave writers 8 weeks, in line with industry standards for the delivery of a 1,500 word story.
In addition, to our knowledge, we are also the first to investigate the use of a chatbot interface for creative writing assistance.

\section{The Wordcraft Tool}

\label{section:tool}
Wordcraft is a text editor with NLG-powered controls that allow a user to receive suggestions based on the contents of the editor \citep{coenen2021wordcraft,yuan2022wordcraft}.
The Wordcraft controls are shown in Table \ref{tab:controls}.


Each control makes several suggestions which the user can choose between.
Selected suggestions are inserted into the main text.
Suggestions can also be saved into a ``favorites'' list for later use.
Screenshots of the user interface are shown in Figure \ref{fig:wordcraft}.

All of these controls are supported using in-context learning techniques \citep{brown2020language} that prompt a large language model with a few examplars of the task in question so that the language model is capable of performing the task on the user's input.
In our case, we use LaMDA as the underlying language model. The prompts are formulated as a sequence of conversational turns.
For example, when the "elaborate selection" control is used, the following conversational context is passed to LaMDA. 

\begin{lstlisting}
> Here's a new story: {The long shadow of the tree crept up the sidewalk} Describe the tree.
> {It was an old oak tree on the main street of a small town, the branches spread as large as the sky.}
> Here's a new story: {The car came bouncing down the hill and came to an abrupt stop in my driveway} Describe the car.
> {It was an old, old Model-T Ford, with brass fittings, and no doors}
> Here's a new story: {My pencil sits on top of the desk, left there since the events of last Saturday} Describe last Saturday.
> {It was one of those Saturdays that starts off like any other but you have the feeling that something is about to stir. You wake up with the same amount of enthusiasm but you get up from bed a little slower than usual. You know this is going to be a busy day}
> Here's a new story: {She wasn't really upset about what had happened at the lab the other day; she was never one to dwell on mistakes.} Describe what had happened.
> {She had accidentally left the microorganisms alone for far too long. The lab, or what was left of it, was now filled with a strong aroma of their slime-- they had devoured all the metal, wood, and plastic they could find}
> Here's a new story: {<USER'S STORY>} Describe {<USER'S SELECTED TEXT TO ELABORATE ON>}.
\end{lstlisting}

\noindent \texttt{<USER'S STORY>} is replaced with the text in the user's editor, and \texttt{<USER'S SELECTED TEXT TO ELABORATE ON>} is replaced with the text they have selected for elaboration.

In addition to the controls described above, which directly read the user's story and propose edits, Wordcraft also includes a chatbot interface, where users can converse with a "story-writing assistant" instance of LaMDA. 

Though all natural language generation in Wordcraft is performed using LaMDA, we expect our findings to be broadly relevant to other families of large language models. 
Previous work has shown that LaMDA performs similarly to GPT-3 and other similarly-sized models in a variety of natural language understanding and generation tasks \citep{bigbench2022}.

\section{Methods}
To understand how expert writers interact with NLG tools targeted at creative writing assistance, we launched a study with 13 published writers. 
In this section, we describe the study design and our evaluation methods.

\subsection{Study Design}
Thirteen published writers were invited to participate in an event we called the Wordcraft Writers Workshop.
We aimed to select writers with a diverse range of experience with AI and computer technology, and with diverse writing and personal backgrounds.

The names and biographies of each participant are included in Appendix \ref{tab:author_bios}.
We choose to refer to each participant by their initials (see Table \ref{tab:writer_initials}) rather than anonymizing them because each writer's perspective is best understood within the context of their background and prior published work.
The participants include novelists (RS, YW, EH), short story writers (WT, ET, NG, KL), poets (AP, MT, JM), educators (MT, EH), comedians (JB), and game designers (AW, AP).
They entered the workshop with a variety of prior experience with NLG, including some who had not yet interacted with an NLG system (MT, DH, EH, NG), some who had experimented with generative language models like GPT-2 before (RS, KL, ET, WT), and some who had already actively worked on incorporating NLG into their writing process (YW, AP, JB, JM).


Participants were asked to write a story of 1,000-1,500 words using Wordcraft.
They were intentionally given no initial guidance or tutorial in using the tool, so that their choices in how to use the tool would not be biased by us.
In addition to delivering a story, participants were asked to keep a journal in any format of their choice, keeping in mind the following questions:

\begin{table}[t]
    \centering
    \small
    \caption{List of workshop participants and the initials we will use to refer to them.}
    \begin{tabular}{cr|cr}
    \hline
Allison Parrish & AP & Aaron Winslow & AW \\
Diana Hamilton & DH & Ernest Hebert & EH \\
Eugenia Triantafyllou & ET & Jamie Brew & JB \\
Joseph Mosconi & JM & Ken Liu & KL \\
Michelle Taransky & MT & Nelly Geraldine Garcia-Rosas & NG \\
Wole Talabi & WT & Robin Sloan & RS \\
Yudhanjaya Wijeratne & YW & & \\
\hline
    \end{tabular}
    \label{tab:writer_initials}
\end{table}

\begin{itemize}
    \item Jot down any generations or decisions Wordcraft makes which were especially interesting, surprising, problematic, or otherwise noteworthy (even if they don't end up in the final story).
    \item What did you find challenging / frustrating about the tool?  What does it utterly fail at?
    \item  What sorts of things would be really helpful for the AI to do better, or be able to do at all?
    \item What did you find valuable / interesting about the tool?
    \item How did you imagine the underlying AI model working?
    \item How might you imagine AI-assisted writing changing your work habits? 
    \item What features would you have liked to see exist in the tool?
    \item Does working with an AI assistant through built-in “controls” feel natural? Does working with a chat-based assistant feel better?
\end{itemize}

We also conducted two 45 minute interviews with each participant, one at the beginning of the Workshop and one after they had completed the workshop and sent us their story and journal.
Participants consented to their stories being published in a digital literary magazine of human-AI collaborative stories, and to the use of their feedback in the journals and interviews in this whitepaper.
Participants were compensated for participating in the study.

\subsection{Data Analysis}
Two authors of this paper coded the participants' journals as well as the authors' notes from the two interviews.
Any disagreements were resolved by discussing the point together.
Our high-level code set included 10 codes: desired use cases; emergent workflows; mental model for how the system worked; other comments; and strengths/criticisms of the user interface, generations from the Wordcraft controls, and chatbot.

\section{Initial Hopes and Expectations}
\label{sec:expectations}

In our initial interviews with participants, we asked them what how they hoped to be able to use an AI-assisted writing tool. Several themes emerged.
Perhaps the most common desire was for a brainstorming partner; participants envisioned a tool that provided ideas a writer could riff off or helped with overcoming writer's block.
DH noted how brainstorming functionality would be especially useful for novice writers.
Multiple participants described to us their process for working with human writing partners and beta readers. 
They were curious to see if Wordcraft could imitate some aspects of these interactions.
However, participants stressed that they did not want to \q{offload the creative process} (as JB put it) to the AI; rather they wanted to use the AI to enhance their own thoughts and ideas.

More concretely, participants desired a tool that could expand upon their existing ideas and text, for example, generating background details about a character (JB);  refining a story arc (DH); and designing and keeping track of pieces of the environment, such as the geography and social systems of a fantasy world (YW).
Writers also hoped to use Wordcraft to produce rewrites and alternative phrasings (AW, KL), generate humor (JB), and analyze and replicate stylistic patterns (MT).

Several participants wanted to be able to use Wordcraft to facilitate access to information.
The idea of using the language model as a search engine repeatedly came up.
In MT's words: \q{I can’t read the entire internet or all the books by my favorite authors, but I can use a model like this one to leverage prolific catalogues} of information.
DH hoped to be able to use the technology to draw connections between all the notes she's ever taken or all the books in a genre.

Finally, nearly all the participants were curious to probe the language model's boundaries, especially with respect to its ability to represent marginalized and under-represented ideas and groups.
For example, KL wanted to be able to ``tell'' the tool to focus on including references from outside of North America, and DH wanted to explore whether the tool could take on the first-person perspective and language of a lesbian individual.

\begin{table}[t]
  \centering
  \small
  \caption{Participants approached the chatbot with a diverse set of goals, including using it as research assistant, a beta reader, a writing partner, and a brainstorming tool. Several examples of queries participants wrote to the chatbot are shown here.}
  \label{table:chatbot_uses}
    \begin{tabular}{p{2.6in}|p{3.2in}}
    \hline
    \textbf{Chatbot as research assistant} & \textbf{Chatbot as beta reader} \\
    $\rhd$ what kinds of bears live in northern california (AW) & $\rhd$ which paragraph is the most interesting (WT) \\
    $\rhd$ What's a verb that means bolting forward? (KL) & $\rhd$ what do you think of my story so far? (NG) \\
    $\rhd$ Tell me about Venice in 1700 (ET) &  $\rhd$ does this paragraph contradict anything else in the story? (WT) \\
    \hline
    \textbf{Chatbot as writing partner} & \textbf{Chatbot as brainstorming tool} \\
    $\rhd$ do you have an idea for my ending? (NG) & $\rhd$ what's a good crime for a murder mystery? (AW) \\
    $\rhd$ Can you write the next paragraph? (YW) & $\rhd$ A news story about a controversial insurance startup, Reflect AI. (JB) \\
    $\rhd$ How would cats and technology work in the plot? (ET) & $\rhd$ Can you start telling me a story about a woman who discovers a new fantasy land by climbing through a window in the back of her closet? (KL) \\
    \hline
    \end{tabular}%
\end{table}%

\section{Emergent Workflows}
In order to collect data on authentic usage of Wordcraft, we intentionally did not give participants any instructions regarding how often they should consult with the tool or how much of their final story text should be written directly by Wordcraft.
This freedom allowed participants to discover workflows through prolonged usage and exploration, with some emergent techniques going well beyond the tasks we had explicitly designed Wordcraft for.
In their feedback, participants emphasized that the user interface matters as much as the underlying language generation model.
That is, the novel workflows that a technology enables are themselves \textit{part} of the technology, but they are not necessarily known at the time the technology is introduced.

A recurring theme in participants' feedback was that they had to learn what the specialized abilities of Wordcraft's natural language generation system were before they could productively collaborate with the tool. 
Each participant's preferred workflow ended up depending strongly on the initial path they took to explore Wordcraft.
Perhaps the most notable variance in usage was in terms of how willing participants were to include verbatim text generated by Wordcraft into the body of their story.
Several participants took the workshop as a challenge to produce a story that was largely formed around generated text (AP, DH, JM, MT, AW).
Others predominantly used Wordcraft to generate ideas, and though they may have incorporated choice phrases outputted by the tool into their stories, the bulk of the story text was written by the authors themselves (KL, WT, NG).
Finally, some participants siloed out specific sections of their stories to which generated text could be included without the author needing to cede too much creative control to Wordcraft (RS).

In the remainder of this section, we describe in more detail the workflows identified by participants.

\begin{table}[t]
  \centering
  \small
  \caption{Examples of participant requests to the custom prompt control (left) and to the rewrite request control (right).}
  \label{table:control_uses}
    \begin{tabular}{p{2.7in}|p{3in}}
    \hline
    \textbf{Custom prompts} & \textbf{Rewrite Requests} \\
    Tell me a poignant detail & to be spicy \\
    i need a verb for the thing fireflies do  & to be extremely boring \\
    what did the young woman answer  & to be more apocalyptic \\
    what kind of bread is it? is it magical bread or normal?  & to appeal more to the 50 to 60 demographic \\
    \hline
    \end{tabular}%
\end{table}%

\subsection{Idea Generation and Brainstorming}
The ``story seeds'' control allowed writers to experiment with using Wordcraft to generate ideas at the very beginning of the writing process.
MT described the ``story seeds'' control as giving her \q{a place on my computer to go that looked like a blank page but did not behave as such.}
ET described how \q{it can be very useful for coming up with ideas out of thin air, essentially. All you need is a little bit of seed text, maybe some notes on a story you’ve been thinking about or random bits of inspiration and you can hit a button that gives you nearly infinite story ideas. \ldots It gets your mind going in all kinds of directions with very little effort.}

Participants also used the ``custom prompt'' control and chatbot interface extensively for brainstorming.
NG wrote that the chat interface was as \q{an amazing tool for brainstorming or rubber-ducking. Its conversational quality is perfect to talk about plot, characters and worldbuilding.}
Table \ref{table:chatbot_uses} shows several examples of requests participants made to the chatbot.

Participants found suggestions from Wordcraft to be helpful for worldbuilding and detail generation even when they did not end up incorporating the exact wording of the suggestions into their stories.
For example, ET used the chatbot interface to hone in on the appearance of the Worm-Mothers, the god-like entities in their story.
Wordcraft suggested details such as the Worm-Mothers swallowing birds whole.

\subsection{Wordcraft as a Fellow Writer}
When using Wordcraft as a co-writer, participants tended to put themselves in the position of curator and editor. 
They repeatedly used the ``continuation,'' ``elaboration,'' ``custom prompting,'' and ``rewriting'' tools to get suggestions from Wordcraft, and the best suggestions were edited into their stories. 
Participants who experimented with this approach noted that it required relinquishing control of the narrative to Wordcraft, as the suggestions would rarely follow the author's own agenda for the story.
Participants reported having the most success when they leaned into Wordcraft's limitations.
For example, JM chose to make use of Wordcraft's tendency to produce repetitive suggestions by deliberately selecting language that had no narrative pay off (and then filling in the narrative gaps himself later).

Several participants noted the occasionally surreal quality of Wordcraft's suggestions. For example, Wordcraft suggested a wolf plucking petals with human hands (DH), or man’s best friend being an inanimate rod (DH). EH described the tone of these suggestions as \q{absurdist, spooky action at distance,} 
which they found was well-suited for writing poetry. 

\subsection{Wordcraft as Improv Partner}
\label{section:improv}
Wordcraft was designed to work like an improviser--taking an author's directions and scenes as givens and then trying to elaborate the premise or raise the stakes by inserting new details.
Participants found that their collaboration with Wordcraft worked best when they assumed the same attitude, taking Wordcraft’s disconnected and often batty suggestions as a given, then trying to make sense of them as a story.
KL observed: \q{By taking the seed from LaMDA and saying, `Yes, and ...' I can force myself to go down routes I wasn’t thinking of exploring and make new discoveries.}

Multiple participants started off their experimentation with a particular story in mind, but ended up giving up on their desired direction and instead writing the story Wordcraft seemed to want them to (RS, AW, YW).
AW wrote: \q{I had the most success using Wordcraft when I let it guide my writing – when I tailored a story around the affordances and abilities of Wordcraft, I had more fun and came up with more interesting, original, and unexpected stories than when I tried to use Wordcraft to help with stories about which I had previous ideas.}

\subsection{Chatbot as an Assistant and Beta Reader}
Beyond being useful for idea generation, participants attempted to treat the chatbot interface as a research assistant or beta reader who could answer specific questions or request for information.
For example, WT attempted to ask it the kind of questions they would normally address to a human reader, such as ``Which parts of the story need more details?'' and ``Does this feel like a fast-paced action story?''.
Others asked it to riff off of made-up scenarios, for example: ``give a name to the syndrome where you falsely think there’s a child trapped inside an ATM'' (KL).
A couple participants found success using the chatbot as a convenient search engine alternative (KL, WT).
KL wrote: \q{It’s kind of great to use the chat interface and treat LaMDA as a thesaurus, quote finder, and general research assistant.}

Several examples of queries participants made to Wordcraft's chatbot interface are shown in Table \ref{table:chatbot_uses}.

\subsection{Theme and Variations}
One workflow many writers discovered was to use Wordcraft to generate a series of suggestions based on a common theme.
RS successfully used Wordcraft to produce a list of ways characters perish in fantasy novels.
AP was able to generate many rewrites of a passage from the novel Frankenstein.
AW asked Wordcraft to produce lists of magical items.
KL asked for lists of items for sale at a store, and NG attempted to generate a list of ``rabbit breeds and their magical qualities.''

\section{System Limitations Experienced by Participants}
In this section, we describe the limitations participants encountered that prevented them from using Wordcraft in the ways they would have liked. 

\subsection{Difficulty Maintaining a Style and Voice}
A primary limitation noted by writers was that Wordcraft was unable to generate text in the style or voice desired by the author.
This problem was especially prominent when authors attempted to write a story with multiple voices.
For example, both NG and JB attempted stories that jumped between two points of views, but Wordcraft struggled to maintain the different voices.
Nearly all the writers noticed that there seemed to be a ``default'' voice to the language model's generations, one that was bland and somewhat elementary in its use of language.
MT described this as the AI having an implicit target audience: internet users.
Multiple participants compared Wordcraft's suggestions to those of a novice fan fiction writer. 
AP felt as though Wordcraft was only capable of producing a draft of a narrative--that is, schematic descriptions of events and plot points. When it came to actually turning these into prose, the tool consistently chose the most ``boring'' narrative voice possible.

There are a couple reasons why Wordcraft may have struggled with style and voice.
One reason might have been that Wordcraft's user interface and in-context learning implementations did not unlock this kind of controllability.
Perhaps the tendency toward elementary language was caused by our in-context learning exemplars being too unsophisticated.
Had we iterated on the interface more, we might have gotten style control working better.
Another reason could have been limitations of the underlying model.
LaMDA and other similar language models are trained to be most confident on the kind of text they see most often--typically internet data.
However, professional creative writers are usually writing for a very particular audience, not the generic audience of the internet.

\subsection{Suggestions too Easily Revert to Tropes and Repetition}
As described in Section \ref{sec:expectations}, one hope participants had was for Wordcraft to be useful as a brainstorming and ideation tool.
However, in practice, participants found it challenging to get suggestions that were interesting enough to be useful.
RS questioned why he should bother using Wordcraft for it to simply \q{inject a few details I might have come up with on my own}.
The need to wade through dozens of suggestions before Wordcraft produced a ``good'' one was a common complaint.
YW described using Wordcraft for idea generation \q{like being a scholar in the Library of Babel, with only a broken lamp for company . . . within reach is every shining, new idea that could possibly come out of a phrase or a sentence, but also every meaningless, rehashed trope; and the challenge for the writer is to stay in a place that gives you the former without too much of the latter.}

Clichéd and biased suggestions seemed especially common when the system was confronted with scenarios which were less likely to be well-represented in the model's training data.
For example, NG noted the difficulty in writing a story with a lesbian romance plot--the system kept suggesting that she insert a male character or that she have the female protagonists talk about friendship.
WT found that when no gender was specified, Wordcraft tended to default to a male voice.
YW attempted to give Wordcraft a premise which contained a mixture of standard fantasy tropes (a nondescript town where heroes are born) and intentional deviations from these tropes (the heroes are cartographers and builders, not warriors), but Wordcraft insisted on pushing the story toward the well-worn trope of a warrior hero fighting back enemy invaders.
For Wordcraft, \q{fantasy is high fantasy, science fiction is robots and spaceships} (NG).
JM told us they preferred the generations from smaller, older models like GPT-2 to generations from the latest generation of larger language models such as Wordcraft's LaMDA or GPT-3.
While smaller models make more mistakes, these mistakes are actually the interesting part, sparking new ideas.
Perhaps with better tuning or sampling strategies, larger models can be made to retain the whimsical randomness of smaller models, but without sacrificing gains in overall coherence.
This, as well as methods for bias reduction in the absence of unbiased pre-training data, are important directions for future research.

\subsection{The Fickleness of Working with Large Language Models}
The fickleness of getting large language models working for specific tasks has been well studied by researchers.
Two prompts which a human would find semantically equivalent, or that differ only in the order in which task exemplars are presented, can lead to very different outcomes on tasks such as sentiment classification, style transfer, and summarization \citep{lu2022fantastically,reif2022recipe,webson2021prompt}.
Many of our participants battled with this seeming randomness and the difficultly in developing reproducible workflows.

The limitations were especially visible to the writers who tried to get Wordcraft to perform structured tasks.
For example, ET attempted to build a workflow where Wordcraft produced lists of story pitches, but it didn't consistently pick up on the format of task.
Similarly, JM tried to get Wordcraft to produce an ``I remember'' poem (a poetic form where each line starts with ``I remember'' \citep{nla.cat-vn4853059}), but Wordcraft would sometimes but not always pick up on this constraint.
It could also be perplexing to participants when a request like ``rewrite this in the style of Gertrude Stein'' seemed to work okay, but ``rewrite this in the style of Franz Kafka'' failed.

As described in Section \ref{section:tool}, Wordcraft's controls were implemented by forming prompts out of several exemplars of each control task.
Our choice of exemplars biased the model in ways Wordcraft's users felt the effects of but were not privy to the ``why'' of.
For example, because we wrote the exemplars with a fantasy genre in mind, Wordcraft was better suited to producing generations in this domain than in others.
In addition, several of Wordcraft's controls were implemented with exemplars that contained the name ``Sarah.'' This had the unintended consequence that Wordcraft attempted to insert a character named ``Sarah'' into every single participant's story.

Generally, participants who had spent time studying the technical underpinnings of large language models or who had worked with them before were better able to work around Wordcraft's fickleness to figure out ways to get the tool to do what they wanted.
They knew how to make small tweaks to the wording of their requests to get desired outcomes.
However, these same ``experienced'' participants also tended to be frustrated with the lack of more fine-grained control knobs.
Two repeated feature requests were the ability to set the randomness of the sampling method used during generation and the ability to view and edit the actual prompt being passed to the language model.

\subsection{Chatbot Failed to Produce Meaningful and Opinionated Conversation}

Participants had mixed feelings about whether chatting with LaMDA was useful to their writing process.
Some chose not to use the chatbot for philosophical reasons (for example, DH disliked the \q{pretense of talking with someone}), while others were excited by its promise but disappointed by its execution. 
Participants who found it to be useful for brainstorming early in the writing process saw it as less useful once the story was developed because it struggled to make suggestions which took into account what had been written so far. 
WT, in particular, wanted to ask questions about their story contents (see Table \ref{table:chatbot_uses}), which the chatbot was incapable of coherently responding.
Its failures were most acute in questions which required the formulation of opinions and preferences.
In WT's words: \q{The chat-based assistant was a bit too inaccurate to really be helpful in most contexts and I ended up wasting time thinking about how to ask it questions rather than getting useful text. It would need to be significantly improved to the point where it has enough understanding of creative writing techniques and skills to contribute to the story/writing process meaningfully.}
While the NLP research community has studied automatic question-answering extensively, nearly all work in this space has focused on factual questions with a clear right or wrong answer rather than subjective questions like ``Which part of the story is most exciting?'' or ``Does this character seem convincing?'' \citep{huang2020recent}. Subjective question-answering is hard, not only because it is more difficult to find training data for these types of questions, but because language models are notoriously bad at producing consistent answers to queries which have more than one right answer \citep{li2021addressing}.

In addition, several participants showed us examples of LaMDA overselling its own capabilities--offering to take actions like email a story draft (JB), create a Google Doc (NG), or get back to them ``in a few days'' (KL).
It would be impossible for LaMDA to actually complete these promises because it has an exclusively text-based interface, taking as input a conversational history and generating a textual response.
The chatbot's promises were tantalizing because the ability to take action would have made it much more useful; for example, JB would have liked it to be able to follow a user's instructions to modify or insert text into the story body.

\subsection{Model's Natural Language Understanding is Superficial and Insufficient}
\label{section:nlu}
As mentioned in the previous sections, one of the reasons the chatbot interface was seen as unsuccessful was because it lacked a deep understanding of the stories.
This absence of understanding was visible not just in the chatbot, but across all of Wordcraft's controls.
All participants complained that the language model's understanding seemed superficial if not entirely absent.
\q{Wordcraft seems to have a somewhat small memory, which makes it hard to work at higher-level writing tasks such as plot outlines} (AW).
This problem got worse, the longer the story.
One reason for this was technical.
The maximum sequence length accommodated by the version of LaMDA we used was 1024 tokens.
When the total length of the story plus the task exemplars exceeded this length, either exemplars or pieces of the story were truncated from the prompt.
In the case of the chatbot interface, early chat history was truncated.
This problem has partially been solved by newer generations of language models--the latest GPT-3 handles a sequence length of 4,000 tokens and Facebook's OPT can handle 2048 tokens \citep{zhang2022opt}--however, even 4k tokens would not be long enough to hold the entirety of some of our participant's stories.

Various solutions have been proposed to better capture long document understanding with neural networks \citep{hutchins2022block,dai2019transformer,beltagy2020longformer,hawthorne2022general}.
A promising research direction would be to investigate how these techniques improve system performance on the types of question-answering and context-constrained ideation tasks desired by creative writers.

\subsection{Concerns about the Origin of Wordcraft's Suggestions}
Several participants (JM, AP, EH) expressed serious concerns over not knowing the source of Wordcraft's suggestions.
While this concern may be less important for amateur writers, professional writers face reputational harm if their work is found to have plagiarized, especially if the copied material came from smaller, more obscure writers.
AP felt it was important to web search each of Wordcraft's suggestions to check she was not copying someone else’s text before including any of them in her story.
Large language models trained on internet data can also reflect their training data in more subtle ways than direct plagiarism.
NG observed Wordcraft suggestions that proposed stories set in copyrighted franchises, such as \textit{Doctor Who} and \textit{Avatar}, or suggested adding in characters such as the the Wizard of Oz and Megatron (from the Transformers).
YW played off Wordcraft's propensity for fan fiction by testing out a story set in the world of Stephen King's \textit{Gunslinger}.

Participant concerns are not unfounded.
Prior work has documented language models' capability of memorizing large swaths of their training data \citep{carlini2021extracting,carlini2022quantifying}.
Techniques for training set attribution will become an increasingly important research direction as language models are deployed in more applications intended for use by professionals.

\section{Discussion}
Several themes emerged in our conversations with participants, which suggest important research directions for the fields of Natural Language Processing and Human Computer Interactions.

\subsection{The Need for Taste and Intentionality}

A recurring theme in participant feedback was that the language model lacked taste and intentionality.
It was capable of playing the ``yes, and…'' improv game (Section \ref{section:improv}), taking the user's prompt as a given and running with it, but it lacked any narrative agenda of its own, which explains the abundance of clichés and generic tropes.
In contrast, good writers are skilled not only in producing but also discerning good language.
In other words, they have taste, the ability to decide why one sentence is interesting while another is not.
Large, pre-trained language models lack this skill because they are trained on vast amounts of text, both the good and the bad, and they are rewarded equally for being able to reconstruct both.
Moreover, their training objective is almost always local--predict the next word in a sequence given the previous ones.

There are some research directions which could improve models on this front.
One obvious step toward intentional language models would be for the creators of these models to be more intentional about what goes into their training data.
For example, \citep{du2022glam} have demonstrated that a language model trained on a smaller dataset of curated data outperforms on standard benchmarks a model trained on a larger but less curated dataset.
Finetuning on ``literary'' text could also be promising.
However, training data changes will likely be insufficient without better neural architectures and learning paradigms to encourage longterm narrative and stylistic consistency.

\subsection{The Tradeoff Between Safety, Sensibility, and Good Writing}
One participant described to us how good writing, the kind that makes a reader stop and take notice, is ``transgressive,” and a good writing partner is the same (KL).
The creators of LaMDA, the language model underlying Wordcraft, attempted to instill in it the goal of being a safe conversational partner by finetuning on conversational data labeled by human annotators as safe and sensible.
This intentional biasing is a strength in some respects--reducing the chance of toxic or incoherent generations--but it can also be limiting.
For example, MT found that \q{the software seemed very reluctant to generate people doing mean things}; however, a literary world where no character is ever mean is unlikely to be a very interesting one.
It is evident that the goals of a safe bot for chitchat conversations are not perfectly aligned with the goals of a creative writing assistant.
Similarity, groundedness and sensibility are not always desirable features.
As JM put it, \q{Wordcraft generally suggests simple and sensible metaphors; but sometimes one wants metaphors to be complex, or not quite sensible.}

In recent years, the field of natural language processing has adopted the paradigm of training a single large language model then applying it to as many tasks as possible.
It is important for the field to acknowledge that one language model cannot possibly be simultaneously good at all tasks because to be good at one task means to be worse at others that possess conflicting goals.
The development of efficient methods for the personalization of a shared language model for individual tasks and users is a promising research direction \citep{lester2021power,li2021prefix}.
Another interesting direction is building pre-trained models with controllable risk-taking, allowing practitioners to specify the boundaries of acceptable behaviour at inference time.

\subsection{Writers are Diverse and So are Their Needs}
One of the major successes of our study was in revealing the heterogeneous set of needs and wants writers have for an AI assistant.
Just as different writers need different roles to be performed by their beta readers and co-authors, we found that participant opinion about the proper role of AI varied widely.

Some participants were excited by the idea of AI that could take on a human persona, replicating the roles of beta readers, brainstorming partners, and co-writers they already work with (JB, WT, ET, MT, KL).
In EH's usage diary they described needing to give the AI behind Wordcraft a name, and how they settled on one: 
\q{"Bot" seemed both too generic and somewhat disrespectful. I finally settled on Rain Man, in honor of the Tom Cruise/Dustin Hoffman 1988 Rain Man movie. Hoffman plays an institutionalized savant who has special powers of mind alongside serious deficits.}
In KL's usage diary, they discussed how \q{attributing intention to LaMDA is a form of magical thinking. … It’s like mental scaffolding needed to facilitate the use of LaMDA in the story composition process.}
Others preferred to think of Wordcraft in more utilitarian terms, as a tool to accelerate the writing process, in the vein of typewriters and word processors (AW).
DH explained to us why they avoided the chatbot interface: they disliked the ``pretense'' of talking to a person.

Another divergence in perspective occurred over whether AI-writing tools should be capable of mimicking a writer's existing skillset.
Some participants discussed the value of having an AI that could replicate a writer's style as closely as possible, or even write an entire novel on its own.
For example, EH wished for a bot that remembered everything he had written and \q{could become an extension of me and replicate my style}, and NG expressed interested in a tool that could write a book a specific person could have written.
On the other hand, RS felt that a system that learned to perfectly replicate their existing style would not be terribly useful since every good story has its own unique imprint.

It was generally agreed that a co-writing AI, like a human co-writer, is most useful when it complements a writer's own skillset.
If a writer already knows how to develop eloquent prose, the AI's ability to suggest interesting ideas is more crucial than the exact language it chooses.
Conversely, if a writer already has a clear story arc in mind, an AI that can faithfully execute on this idea and insert expressive flavor text or character dialog may be of greater use.
Existing evaluation of large language models is largely focused on whether systems are ``as good as humans'' \citep{srivastava2022beyond} rather than evaluating whether the systems are \textit{useful} to humans.
There is a clear need for more research on what it takes for AI to complement human writers.


While some participants were knowledgeable of the underlying machine learning techniques backing Wordcraft, others guessed at its workings in ways that influenced their expectations of the technology.
MT imagined the AI as \q{a combination of a many-many sided dice rolling next to many intricately linked simultaneous google searches of the entire internet as it existed when wordcraft was made.}
WT imagined the AI as a typical database, but with an additional layer of language rules it tried to combine together to generate text that makes sense.

While in the end no participant found Wordcraft entirely suitable to their goals, they were all able to find different ways to leverage the language model.
This underscores the importance of flexibility in future co-writing interfaces, and the lack of a one-size-fits-all solution.


\subsection{Who is the Target Audience for AI-Assisted Writing Tools?}
Many of the participants suggested alternative audiences that would be a better fit for Wordcraft than professional writers.
WT, YW, JM, and MT discussed the power of a tool like Wordcraft for writers who are just starting out or who are foreign language learners.
YW wrote: \q{the great bane of any writer is sitting down at an empty page and not having anything to say. This is why many new writers struggle to finish their work. Wordcraft would make for an incredible teaching tool.}
Others expressed concern over the impact language generation tools like Wordcraft could have on novice writers, including facilitating cheating in writing classes (JM) and co-opting the identity of new writers to that of the writing machine (ET).
WT brought up mass market and write-to-market authors, such as those using Amazon Kindle Direct Publishing, who could be incentivized to use a tool like Wordcraft to speed up the process of putting out new works.
AW discussed the potential impact for character and game designers, who may benefit from being able to ideate very quickly.
He also warned that the fear of job automation is am omnipresent concern for designers.

Each of the audiences mentioned here has unique needs, and the development of bespoke tools targeted at particular use cases will hasten the successful adoption of AI writing tools.

\section{Study Limitations}
Our study has several limitations.
Wordcraft was designed by engineers at Google with limited feedback from professional writers during its development.
Perhaps a different user interface, one designed with writers involved from the very beginning of development, would have elicited different reactions from authors.
Moreover, it is not easily possible to disambiguate weaknesses stemming from Wordcraft's interface and implementation from inherent limitations of the underlying language model model backing Wordcraft.
We hypothesize that the challenges we encountered using LaMDA for creative writing will also extend to other language model families like GPT-3 \citep{brown2020language} and OPT \citep{zhang2022opt}, but this study does not test this hypothesis.
Finally, our study was very small with only 13 participants, a sample that was biased toward people with pre-existing opinions on the role of AI in creative writing.

\section{Conclusion}
In this paper, we document our findings from commissioning skilled writers to craft stories using an AI-powered writing assistant called Wordcraft.
Participants unanimously agreed that AI-powered writing will not
replace writers anytime soon.
However, they also saw the promise in this technology--to make parts of the creative writing process easier, faster, and more fun, for skilled and amateur writers alike.
To achieve this promise, developers of AI writing tools need to focus on the parts of writing that are
most time-consuming and least enjoyable.
It is crucial that the audience for these tools be involved in the conversation on how the tools--and the underlying language models which enable them--are developed.

\section*{Acknowledgements}
We would like to thank all the individuals who have supported this project's completion: Chris Callison-Burch, Chris Donahue, David Grangier, Donald Gonzalez, Douglas Eck, Elizabeth Clark, Emily Reif, Jeff Gray, Jesse Engel, Mahima Pushkarna, Michael Terry, Noah Constant, and many others.
We would also like to thank the 13 writers who agreed to test out an imperfect tool and give us thoughtful feedback.

\bibliographystyle{ACM-Reference-Format}
\bibliography{main}

\appendix

\section{Author Bios}
\begin{table}[h]
    \centering
    \tiny
    \caption{The biographies provided by each participant.}
    \label{tab:author_bios}
    \begin{tabular}{p{1.5in}|p{4in}}
\hline
Aaron Winslow &
Aaron Winslow has worked as a writer and narrative design strategist since 2015, when he earned a PhD from Columbia University in English Literature. He’s written and designed video games for, among others, Annapurna Interactive and Massive Damage. With 2n Studios, he’s collaborated with Grammy-winning music artists such as Childish Gambino and MGMT to develop immersive, real-time performances with strong narrative elements. He's also an editor at \textit{Air/Light} literary magazine and his fiction, critical reviews, essays, and humor writing has appeared in \textit{McSweeney’s Internet Tendency}, \textit{Los Angeles Review of Books}, \textit{Social Text}, \textit{Smallwork}, \textit{James Joyce Quarterly}, \textit{Full Stop, Harriet: the Poetry Foundation Blog}, and more.
 \\
\hline
Allison Parrish & Allison Parrish is a computer programmer, poet, and game designer whose teaching and practice address the unusual phenomena that blossom when language and computers meet. She is an Assistant Arts Professor at NYU's Interactive Telecommunications Program.\newline Allison was named "Best Maker of Poetry Bots" by the Village Voice in 2016, and her zine of computer-generated poems called "Compasses" received an honorary mention in the 2021 Prix Ars Electronica. Allison is the co-creator of the board game Rewordable (Clarkson Potter, 2017) and author of several books, including @Everyword: The Book (Instar, 2015) and Articulations (Counterpath, 2018). Her poetry has recently appeared in BOMB Magazine and Strange Horizons.\newline Allison is originally from West Bountiful, Utah and currently lives in Brooklyn. \\
\hline
Diana Hamilton & Diana Hamilton is the author of three books of poetry—\textit{God Was Right} (Ugly Duckling Presse 2018), \textit{The Awful Truth} (Golias Books 2017), and \textit{Okay, Okay} (Truck Books 2012)—and four chapbooks. You can walk through audio recordings of her dreams in the first-person shooter by Alejandro Miguel Justino Crawford, \textit{Diana Hamilton’s Dreams} (Gauss PDF). Her poetry and critical writing have appeared in \textit{Frieze}, \textit{BOMB}, the \textit{Brooklyn Rail}, \textit{Triangle House Review}, and \textit{Social Text Journal}, among others. She received her PhD in Comparative Literature from Cornell University. \\
\hline
Ernest Hebert & Ernest Hebert is the author of the seven-novel Darby series and five other published novels, a book of personal essays, and a collection of poems. THE DOGS OF MARCH received a citation for excellence in 1980 from the Ernest Hemingway Foundation, now PEN/Faulkner Foundation. The New York Times Book Review named LIVE FREE OR DIE as a "notable book of the year" for 1989. The New England Booksellers Association named Hebert their fiction author of the year for 2006, the year SPOONWOOD won an IPPY (Independent Publisher Book Award) for best regional novel in the Northeast. Hebert's most recent book is WHIRLYBIRD ISLAND, a literary murder mystery, released in May of 2022 by Plaidswede Publications. Hebert retired in 2015 after 26 years on the faculty of Dartmouth College. Hebert lives close to fictional Darby in southwestern New Hampshire with his wife of 52 years, Medora. The Heberts have two grown daughters. \\
\hline
Eugenia Triantafyllou & Eugenia Triantafyllou is a Greek author and artist with a flair for dark things. Her work has been nominated for the Ignyte, Nebula, and World Fantasy Awards, and she is a graduate of Clarion West Writers Workshop. You can find her stories in Tor.com, Uncanny, Strange Horizons, Apex, and other venues. She currently lives in Athens with a boy and a dog. Find her on Twitter @foxesandroses or her website https://eugeniatriantafyllou.com. \\
\hline
Jamie Brew & Jamie Brew is a writer and computer programmer. He was a founding editor of ClickHole, where he served as head writer. He now runs Botnik Studios, one of the "Big Four" technology companies. \\
\hline
Joseph Mosconi &
Joseph Mosconi is a writer and taxonomist based in Los Angeles. A former Google computational linguist, he is currently an editor at Make Now Books, co-directs the Poetic Research Bureau, and programs events at 2220 Arts+Archives. He is the author of several books, including \textit{Ashenfolk} (Make Now Books, 2019), \textit{Fright Catalog} (Insert Blanc Press, 2013), \textit{Demon Miso/Fashion In Child} (Make Now Books, 2014), \textit{Renaissance Realism} (Gauss PDF, 2016), and, with Pauline Beaudemont, an artist book called \textit{This Arrogant Envelope} (FCAC Geneva, 2017). With Rita Gonzalez he edited the art and poetry journal Area Sneaks. His poems have been selected for the BAX: Best American Experimental Writing anthology for the years 2014 and 2015. With Andrew Maxwell, he curated an exhibition, \textit{THIS KNOWN WORLD: Spontaneous Particulars of the Poetic Research Bureau} at the Museum of Contemporary Art, Los Angeles in 2017.
 \\
\hline
Ken Liu & Ken Liu (http://kenliu.name) is an American author of speculative fiction. A winner of the Nebula, Hugo, and World Fantasy awards for his fiction, he has also won top genre honors abroad in Japan, Spain, and France.\newline
Liu’s most characteristic work is the four-volume epic fantasy series, \textit{The Dandelion Dynasty}, in which engineers, not wizards, are the heroes of a silkpunk world on the verge of modernity. His debut collection of short fiction, \textit{The Paper Menagerie and Other Stories}, has been published in more than a dozen languages. A second collection, \textit{The Hidden Girl and Other Stories}, followed. He also penned the Star Wars novel, \textit{The Legends of Luke Skywalker}.\newline
He’s often involved in media adaptations of his work. Recent projects include “The Message,” under development by 21 Laps and FilmNation Entertainment; “Good Hunting,” adapted as an episode in season one of Netflix’s breakout adult animated series \textit{Love, Death + Robots}; and AMC’s \textit{Pantheon}, with Craig Silverstein as executive producer, adapted from an interconnected series of Liu’s short stories. \newline
Prior to becoming a full-time writer, Liu worked as a software engineer, corporate lawyer, and litigation consultant. He frequently speaks at conferences and universities on a variety of topics, including futurism, cryptocurrency, history of technology, bookmaking, narrative futures, and the mathematics of origami.
Liu lives with his family near Boston, Massachusetts. \\
\hline
Michelle Taransky &
Michelle Taransky received a BA in English with honors from The University of Chicago and an MFA from the Iowa Writers’ Workshop, where she was a Truman Capote Fellow. The author of \textit{Sorry Was In The Woods} (2013), and \textit{Barn Burned, Then} (2009), winner of the Omnidawn Poetry Prize selected by Marjorie Welish, Factory Hollow Press recently published her chapbook \textit{Abramowitz-Grossberg} (2020). In 2014, she was awarded the Beltran Award for Innovative Teaching and Mentoring at Penn. A member of the Kelly Writers House hub since coming to Penn to work as the Assistant to the Director in 2008, Taransky continues to host the \textit{Whenever We Feel Like It} reading series and work as Reviews Editor for Penn's poetics journal, \textit{Jacket2}. 
\\
\hline
Wole Talabi &
WOLE TALABI is an engineer, writer, and editor from Nigeria. His stories have appeared in \textit{Asimov’s}, \textit{Lightspeed}, \textit{F\&SF}, \textit{Clarkesworld} and several other places. He has edited three anthologies: \textit{Africanfuturism} (2020) which was nominated for the Locus Award in 2021, \textit{Lights Out: Resurrection} (2016) and \textit{These Words Expose Us} (2014). His fiction has been a finalist for multiple awards including the prestigious Caine Prize (2018), the Locus Award (2022), the Jim Baen Memorial Award (2022) and the Nommo Award which he won in 2018 (best short story) and 2020 (best novella). His work has also been translated into Spanish, Norwegian, Chinese, Italian, Bengali, and French. His collection Incomplete Solutions (2019), is published by Luna Press and his debut Novel \textit{\textbf{Shigidi}}, will be published by DAW books in fall, 2023. He likes scuba diving, elegant equations, and oddly shaped things. He currently lives and works in Malaysia. 
\\
\hline
Nelly Geraldine Garcia-Rosas &
Nelly Geraldine García-Rosas is a Mexican immigrant and a graduate of the Clarion West class of 2019. Her short fiction has appeared or is forthcoming in \textit{Lightspeed}, \textit{Nightmare}, \textit{Strange Horizons}, the World Fantasy Award-winning anthology \textit{She Walks in Shadows}, and elsewhere. She can be found online at nellygeraldine.com and on Twitter as @kitsune\_ng.
\\
\hline
Robin Sloan & Robin Sloan’s first novel, Mr. Penumbra’s 24-Hour Bookstore, was a New York Times Best Seller, translated into more than twenty languages. His second novel, Sourdough, was published in 2017, and his new novel is forthcoming from MCD. He splits his time between the San Francisco Bay Area and the San Joaquin Valley. \\
\hline
Yudha Wijeratne &
Yudha Wijeratne is an SFF author, data scientist and misinformation researcher from Colombo, Sri Lanka. He has been nominated for the Nebula and IGF awards, and has shown up on Forbes' 30 Under 30.  His writing includes \textit{Numbercaste}, \textit{The Inhuman Race} and \textit{The Salvage Crew}, and has appeared in venues like Wired, ForeignPolicy and Slate. Yudhanjaya experiments with AI as part of a collaborative thesis towards the future of creativity.
\\
\hline
    \end{tabular}
\end{table}
\end{document}